\documentclass[sigconf,10pt]{acmart}
\usepackage{booktabs}
\usepackage{tabularx}
\usepackage{graphicx}
\usepackage{amsmath}
\usepackage{algorithm}
\usepackage{algorithmic}
\usepackage{pgfplots}
\pgfplotsset{compat=1.18}
\usepackage{enumitem}
\usepackage{tikz}

\usepackage{silence}
\usetikzlibrary{positioning, arrows.meta, calc, fit, backgrounds,
decorations.pathreplacing}

\setcopyright{none}                                                                           
\renewcommand\footnotetextcopyrightpermission[1]{}                                            
\title{Federated Parameter-Efficient Adaptation for Interference Mitigation at the Wireless Edge}

\author{Evar Jones}
\affiliation{
\institution{Virginia Tech National Security Institute}                                     
\country{USA} 
}
\email{jevar@vt.edu}

\author{Daniel J. Jakubisin}
\affiliation{
\institution{Virginia Tech National Security Institute}
\country{USA}
}
\email{djj@vt.edu}

\author{Sanmay Das}
\affiliation{
\institution{Virginia Polytechnic Institute and State University}
\country{USA}
}
\email{sanmay@vt.edu}

\begin{document}
\fancyhead{}
\begin{abstract}
Dense wireless deployments face co-channel interference from heterogeneous sources that vary across base stations (gNBs in 5G). While centralized DNN-based approaches to interference mitigation have shown strong performance, deploying and adapting these models across distributed gNBs via federated learning (FL) requires transmitting full model updates each round, resulting in a cost that scales poorly with network density. Parameter-efficient fine-tuning (PEFT) reduces this burden by training and communicating only a small fraction of parameters. While traditionally applied to large foundation models, we adapt Low-Rank Adaptation (LoRA) to temporal convolutional neural network architectures for interference suppression, placing low-rank adapters on the dilated convolutional layers. This placement enables LoRA to learn local interference-specific temporal patterns, while the frozen backbone retains the shared signal extraction capability. These lightweight adapters (5.1\% of backbone parameters) are federated via FedAvg, reducing per-round communication by up to 20$\times$ compared to federating full model updates. We evaluate various PEFT strategies across simulated distributed gNBs with non-IID interference environments. Results show that local LoRA achieves 12.8\% average BER improvement over the frozen backbone, while Fed-LoRA achieves comparable performance (12.6\%). Fed-LoRA outperforms local adaptation on data-starved nodes where federated knowledge transfer compensates for limited samples, all while avoiding the catastrophic degradation observed with full-model FedAvg under heterogeneous conditions.  

\end{abstract}

\maketitle

\section{Introduction}
The densification of wireless networks toward 6G intensifies co-channel interference from heterogeneous sources, making robust interference suppression at the physical layer an increasingly critical challenge \cite{oyedare2022interference}. Interference suppression is commonly addressed
through signal (source) separation–based mitigation techniques, which aim to recover the signal of interest (SOI) with
high fidelity to support reliable detection, demodulation, and decoding \cite{lancho2025rfchallenge}. Data-driven methods based on deep learning (DL) have demonstrated superior performance for source separation
over traditional methods \cite{tian2024learnablewavenet, kothari2025interference, henneke2024soi}. However, the traditional centralized learning methods may no longer be feasible due to privacy concerns, computational burden, and communication overhead \cite{singh2024communication}. Additionally, the rapid changing wireless environment requires online adaptation \cite{wang2025fedpda}.

Federated learning (FL) is a distributed ML paradigm in which multiple clients collaboratively train a shared model by computing updates locally on their private data and periodically aggregating these updates at a coordinating server, without transferring raw data off the devices. This approach decouples model training from centralized data collection, reducing communication overhead and privacy risks while enabling learning from distributed data sources \cite{mcmahan2017communication}. Various works have explored FL for online adaptation at the physical (PHY)-layer, including for neural receivers \cite{wang2025fedpda} to perform well in different channel environments, channel estimation \cite{elbir2021fl_channel_estimation}, and  modulation classification \cite{rahman2024pfl_modulation}. FL for RF interference suppression remains largely unexplored, despite the task being inherently distributed across base stations that each encounter distinct local interference environments.

Meanwhile, the Open RAN paradigm disaggregates traditionally monolithic, vendor-proprietary RAN equipment into modular components, enabling data-driven network operation \cite{pan2025nextg} and the capability of training ML models \cite{singh2024communication}. This architectural shift exposes per-site signal data at distributed base stations (gNB in 5G) capable of localized learning, creating a natural setting for federated learning across the RAN \cite{brik2025explainable}. However, gNBs may have limited resources \cite{naseri2025highthroughput}, making full model fine-tuning and transmission costly at the wireless edge. Federating all model parameters each round strains bandwidth-limited backhaul links, while local backpropagation through the entire model demands compute and memory that may exceed gNB capacity.

Parameter-efficient fine-tuning (PEFT) addresses both bottlenecks: by training and communicating only a small set of adapter parameters, it reduces local training cost and per-round communication simultaneously \cite{han2024peftsurvey}. Among PEFT methods, Low-Rank Adaptation (LoRA) \cite{hu2022lora} is  particularly well-suited to this setting. LoRA augments frozen pretrained weights with low-rank decompositions, training only compact adapter matrices that can be transmitted directly between gNBs and the aggregation server without reconstruction or post-hoc compression. There is a lack of lightweight adaptation strategies for federated interference suppression at the wireless edge that simultaneously reduces local training cost and communication overhead while preserving the backbone's learned signal representations.

In this work, we apply LoRA to the dilated convolutional layers of a WaveNet-based signal separation model, and evaluate its integration within a FL framework in the context of O-RAN systems for interference suppression. In our framework, distributed gNBs act as federated clients that collaboratively adapt a shared signal separation backbone by training and exchanging only lightweight LoRA adapters, orchestrated by a central aggregation server. By restricting fine-tuning to low-rank modules on the interference-sensitive dilated convolutional layers, our approach simultaneously reduces local training cost, memory footprint, and per-round communication. This approach enables efficient federated adaptation across resource-constrained gNBs.

The contributions of this paper are summarized as follows:
\begin{enumerate}[leftmargin=*,itemsep=1pt,topsep=2pt]
    \item We apply LoRA to dilated convolutional layers of a Wave\-Net-based                            
  signal separation backbone, enabling node-specific adaptation
  of temporal interference filtering while training and communicating                           
  only 5.1\% of the model parameters, yielding a~20$\times$ reduction
  in per-round communication cost compared to full-model federated
  averaging.
    \item We provide an empirical study across $K$ number of gNBs with heterogeneous interference environments under two non-IID regimes: (1) balanced, where all gNBs have
  equal data volume but heterogeneous interference types, and (2) imbalanced,
  where data-scarce nodes have limited exposure to rare interference sources. We show    
  that local LoRA applied to the dilated convolutional layers captures 90\% of full fine-tuning gains, while federated LoRA  
  outperforms local adaptation on data-starved nodes and unseen interference, all while avoiding the 
  catastrophic degradation observed with FedAvg.
\end{enumerate}

\section{Background and Motivation}

\subsection{DL-based RF Signal Separation}
Traditional interference suppression techniques rely on domain knowledge of   
the interference characteristics, matched filtering, and adaptive methods such
as successive interference cancellation (SIC), which face limitations in     
adaptability and scalability in dynamic RF environments                       
\cite{oyedare2022interference}. Deep learning approaches have emerged as      
model-free alternatives that learn interference characteristics directly from 
data, achieving superior performance across diverse interference conditions. The ICASSP 2024 RF Challenge \cite{lancho2025rfchallenge} established a
benchmark for data-driven single-channel signal separation, demonstrating that deep learning architectures such as WaveNet and U-Net achieve
orders-of-magnitude improvement in MSE and BER over traditional baselines
including matched filtering and LMMSE across multiple interference types.

Tian et al.~\cite{tian2024learnablewavenet} introduced learnable dilation parameters into the WaveNet architecture to adaptively modulate the receptive field for RF signal processing, achieving optimal signal separation. Naseri et al.~\cite{naseri2024unet} developed a U-Net operating in the
time-frequency domain via STFT, exploiting OFDM signal grid structure for a
63\% average MSE improvement. Henneke \cite{henneke2024soi} extended the
WaveNet baseline with SOI-matched autoencoders that learn the demodulation
and re-synthetization of the signal of interest. The WaveNet architecture,
originally proposed for speech generation \cite{oord2016wavenet} and later
adapted for speech denoising \cite{rethage2018wavenet}, employs dilated causal
convolutions with exponentially growing receptive fields, enabling efficient
multi-scale temporal processing well-suited to RF waveforms. Naseri et al.~\cite{naseri2025highthroughput} explored architectural
compression techniques including depthwise separable convolutions and
quantization to enable high-throughput interference cancellation on edge
devices, demonstrating the need for computationally efficient signal
separation models in resource-constrained deployments. 

However, these approaches are trained and deployed centrally, assuming access
to all interference types during training. In practice, gNBs may encounter
heterogeneous and location-specific interference environments, motivating
distributed fine-tuning strategies that can tailor a pre-trained backbone to
local conditions.

\vspace{-6pt} 
\subsection{FL in O-RAN}
Federated Learning (FL), introduced by McMahan et al.~\cite{mcmahan2017communication}, enables distributed model training by exchanging model updates rather than raw data, addressing both privacy and bandwidth concerns. The O-RAN architecture disaggregates the radio access network into modular components, including the O-RU, O-DU, and O-CU. These components are managed by near-real-time and non-real-time RICs that enable data-driven optimization through open interfaces \cite{brik2025explainable}. In the O-RAN architecture, where disaggregated gNBs are managed by the RICs over bandwidth-limited backhaul links, FL has emerged as a natural paradigm for collaborative intelligence at the wireless edge. Recent works have applied FL to diverse O-RAN tasks including resource allocation and network slicing \cite{zhang2022federated}, anomaly detection within digital twin security architectures \cite{rumesh2024fl_oran_anomaly}, and joint task offloading with fronthaul routing \cite{ndikumana2023federated}.

However, deploying FL efficiently in O-RAN remains challenging. Singh and Nguyen \cite{singh2024communication} demonstrated that the stringent latency requirements and limited compute resources of RICs necessitate communication-efficient FL methods, proposing compressed gradient techniques to reduce backhaul overhead. More fundamentally, Zhao et al.~\cite{zhao2022federated} show that non-IID data distributions across wireless nodes cause severe performance degradation in standard FedAvg, proposing federated data sharing and weight divergence bounds to mitigate this effect. In our setting, non-IID heterogeneity arises not from label distribution skew but from physically distinct interference environments across gNBs. Additionally, FL for interference suppression in an O-RAN context is largely unexplored, despite it being critical for downstream PHY-layer tasks.

\subsection{Parameter-Efficient Fine-Tuning}

Parameter-Efficient Fine-Tuning (PEFT) methods adapt pre-trained models to downstream tasks by training only a small subset of parameters while keeping
the backbone frozen \cite{han2024peftsurvey}. Houlsby et al.~\cite{houlsby2019adapters} introduced adapter modules for NLP, demonstrating near full fine-tuning performance with 3.6\% of the parameters. LoRA \cite{hu2022lora} further simplifies adaptation by injecting trainable low-rank decompositions into frozen weight matrices, avoiding additional inference latency. FiLM \cite{perez2018film} provided an even lighter alternative through channel-wise affine transformations, and has been applied
to federated few-shot learning with fewer than 1\% of updateable parameters \cite{shysheya2023fit}. 

While PEFT originated in NLP and vision transformers, recent work has extended LoRA to convolutional architectures. Conv-LoRA \cite{zhong2024convlora} integrates convolutional experts into LoRA's bottleneck structure for image segmentation, and LoRA-C \cite{ding2024lorac} proposes layer-wise low-rank decomposition for CNN fine-tuning on resource-constrained IoT devices. However, these methods target spatial convolutional architectures for vision tasks. Dilated convolutional networks, which capture temporal patterns across exponentially growing time scales, require adapters that preserve the per-layer dilation structure. This is a design consideration absent from prior work and essential for RF interference suppression.

In federated settings, LoRA's compact parameterization naturally reduces communication cost, as only the low-rank adapter matrices are transmitted rather than full model updates. Several works have proposed refinements to federated LoRA. FFA-LoRA \cite{sun2024ffalora} freezes the A matrices to simplify aggregation and halve communication cost, while CoCo-Fed \cite{guo2025cocofed} compresses full model updates via orthogonal subspace superposition. FAH-QLoRA~\cite{gao2025fahqlora} dynamically adapts LoRA ranks across FL rounds and assigns heterogeneous ranks to devices, reducing training time by up to 45\% on LLM fine-tuning
tasks. In this work, we show that standard federated averaging of LoRA parameters is sufficient for our setting, achieving strong performance without specialized aggregation or compression techniques. To date, federated LoRA has been studied primarily for large language models and vision transformers. No prior work has applied federated LoRA to dilated convolutional architectures or to RF signal separation, where the adapter design must preserve the per-layer dilation structure critical for multi-scale temporal filtering.

\section{System Design}

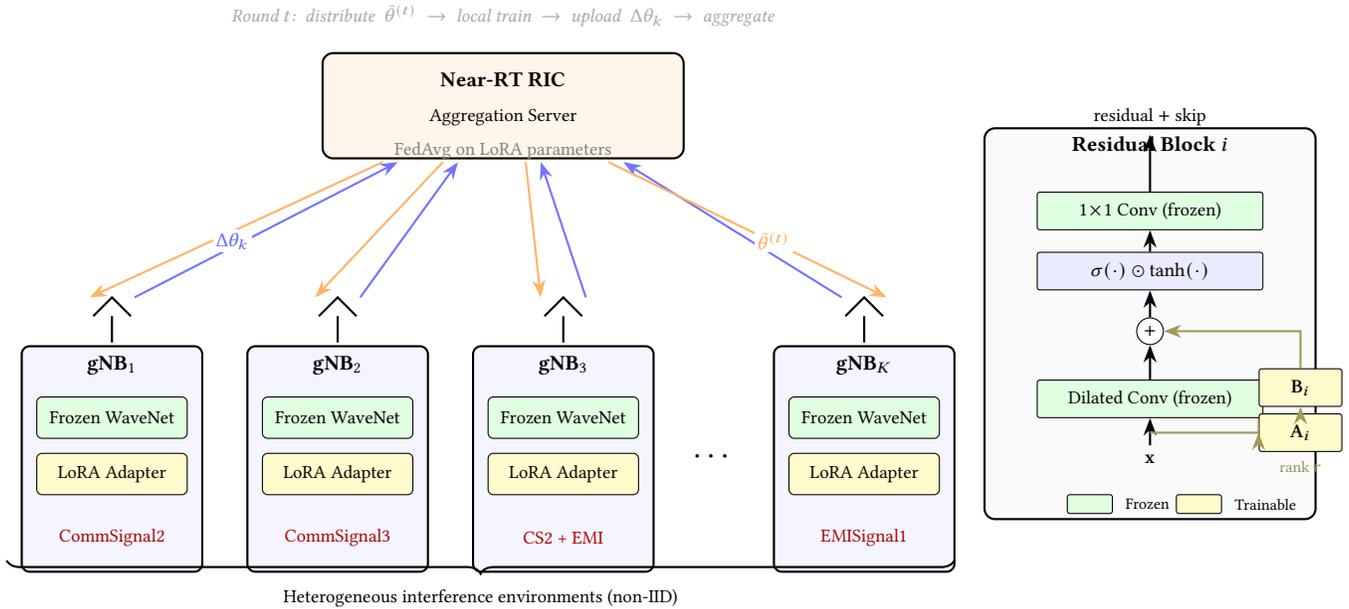
\begin{figure*}[t]                                                            
\centering                                                                    
\begin{tikzpicture}[                                                          
  >=Stealth,
  every node/.style={font=\small},                                          
  gNBbox/.style={draw, thick, rounded corners=3pt, minimum width=2.4cm,
                  minimum height=3.0cm, fill=blue!4},
  modblock/.style={draw, rounded corners=2pt, minimum height=0.55cm,
                   minimum width=2.0cm, align=center, font=\scriptsize,
fill=#1},
  modblock/.default=white,
  wblock/.style={draw, rounded corners=1pt, minimum height=0.5cm,
                 minimum width=2.2cm, align=center, font=\scriptsize,
fill=#1},
  wblock/.default=white,
]

\node[draw, thick, rounded corners=4pt, fill=orange!8,
    minimum width=4.8cm, minimum height=1.4cm, align=center]
  (ric) at (0, 5.5) {};
\node[font=\footnotesize\bfseries] at (0, 5.85) {Near-RT RIC};
\node[font=\scriptsize] at (0, 5.35) {Aggregation Server};
\node[font=\scriptsize, text=gray] at (0, 4.9) {FedAvg on LoRA parameters};


\node[gNBbox] (g1) at (-5.2, 0.8) {};
\node[font=\footnotesize\bfseries] at (-5.2, 2.05) {gNB$_1$};
\node[modblock=green!12] at (-5.2, 1.35) {Frozen WaveNet};
\node[modblock=yellow!25] at (-5.2, 0.6)  {LoRA Adapter};
\node[font=\scriptsize, text=red!65!black] at (-5.2, -0.25) {CommSignal2};
\draw[thick] (-5.2, 2.35) -- (-5.2, 2.7);
\draw[thick] (-5.45, 2.7) -- (-5.2, 2.95) -- (-4.95, 2.7);

\node[gNBbox] (g2) at (-2.2, 0.8) {};
\node[font=\footnotesize\bfseries] at (-2.2, 2.05) {gNB$_2$};
\node[modblock=green!12] at (-2.2, 1.35) {Frozen WaveNet};
\node[modblock=yellow!25] at (-2.2, 0.6)  {LoRA Adapter};
\node[font=\scriptsize, text=red!65!black] at (-2.2, -0.25) {CommSignal3};
\draw[thick] (-2.2, 2.35) -- (-2.2, 2.7);
\draw[thick] (-2.45, 2.7) -- (-2.2, 2.95) -- (-1.95, 2.7);

\node[gNBbox] (g3) at (0.8, 0.8) {};
\node[font=\footnotesize\bfseries] at (0.8, 2.05) {gNB$_3$};
\node[modblock=green!12] at (0.8, 1.35) {Frozen WaveNet};
\node[modblock=yellow!25] at (0.8, 0.6)  {LoRA Adapter};
\node[font=\scriptsize, text=red!65!black] at (0.8, -0.25) {CS2 + EMI};
\draw[thick] (0.8, 2.35) -- (0.8, 2.7);
\draw[thick] (0.55, 2.7) -- (0.8, 2.95) -- (1.05, 2.7);

\node[font=\Large] at (2.8, 0.8) {$\cdots$};

\node[gNBbox] (gK) at (4.8, 0.8) {};
\node[font=\footnotesize\bfseries] at (4.8, 2.05) {gNB$_K$};
\node[modblock=green!12] at (4.8, 1.35) {Frozen WaveNet};
\node[modblock=yellow!25] at (4.8, 0.6)  {LoRA Adapter};
\node[font=\scriptsize, text=red!65!black] at (4.8, -0.25) {EMISignal1};
\draw[thick] (4.8, 2.35) -- (4.8, 2.7);
\draw[thick] (4.55, 2.7) -- (4.8, 2.95) -- (5.05, 2.7);


\draw[->, thick, blue!55] (-4.9, 2.95) -- (-1.4, 4.75);
\draw[->, thick, blue!55] (-1.9, 2.95) -- (-0.6, 4.75);
\draw[->, thick, blue!55] (1.1, 2.95) -- (0.5, 4.75);
\draw[->, thick, blue!55] (4.5, 2.95) -- (1.6, 4.75);

\draw[->, thick, orange!60] (-1.6, 4.75) -- (-5.5, 2.95);
\draw[->, thick, orange!60] (-0.8, 4.75) -- (-2.5, 2.95);
\draw[->, thick, orange!60] (0.3, 4.75) -- (0.5, 2.95);
\draw[->, thick, orange!60] (1.4, 4.75) -- (5.1, 2.95);

\node[font=\scriptsize, text=blue!65, fill=white, inner sep=1pt]
  at (-3.6, 3.7) {$\Delta\theta_k$};
\node[font=\scriptsize, text=orange!70, fill=white, inner sep=1pt]
  at (3.6, 3.7) {$\bar{\theta}^{(t)}$};

\draw[decorate, decoration={brace, amplitude=5pt, mirror}, thick]
  (-6.6, -0.55) -- (6.0, -0.55);
\node[font=\scriptsize, align=center] at (-0.3, -1.05)
  {Heterogeneous interference environments (non-IID)};

\begin{scope}[shift={(8.6, 1.4)}]

  \node[draw, thick, rounded corners=4pt, fill=gray!4,
        minimum width=4.4cm, minimum height=5.2cm] at (0, 1.2) {};
  \node[font=\footnotesize\bfseries] at (0, 3.6) {Residual Block $i$};

  \node[font=\scriptsize] (xi) at (0, -0.6) {$\mathbf{x}$};

  \node[wblock=green!12, minimum width=3.0cm] (dc) at (0, 0.2)
      {Dilated Conv (frozen)};

  \node[wblock=yellow!25, minimum width=1.1cm] (la) at (2.0, -0.25)
      {$\mathbf{A}_i$};
  \node[wblock=yellow!25, minimum width=1.1cm] (lb) at (2.0, 0.35)
      {$\mathbf{B}_i$};

  \node[draw, circle, inner sep=1.5pt, fill=white,
font=\scriptsize\bfseries]
      (sum) at (0, 1.1) {$+$};

  \node[wblock=blue!8, minimum width=3.0cm] (gate) at (0, 1.9)
      {$\sigma(\cdot) \odot \tanh(\cdot)$};

  \node[wblock=green!12, minimum width=3.0cm] (proj) at (0, 2.7)
      {$1\!\times\!1$ Conv (frozen)};

  \node[font=\scriptsize] (yo) at (0, 3.95) {residual $+$ skip};

  \draw[->, thick] (xi) -- (dc);
  \draw[->, thick] (dc) -- (sum);
  \draw[->, thick] (sum) -- (gate);
  \draw[->, thick] (gate) -- (proj);
  \draw[->, thick] (proj) -- (yo);

  \draw[->, thick, yellow!55!black] (0, -0.25) -| (la.west);
  \draw[->, thick, yellow!55!black] (la) -- (lb);
  \draw[->, thick, yellow!55!black] (lb.north) |- (sum.east);

  \node[font=\tiny, text=yellow!50!black] at (2.0, -0.7) {rank $r$};

  \node[wblock=green!12, minimum width=0.6cm, minimum height=0.25cm]
      at (-0.8, -1.2) {};
  \node[font=\tiny, anchor=west] at (-0.45, -1.2) {Frozen};
  \node[wblock=yellow!25, minimum width=0.6cm, minimum height=0.25cm]
      at (0.65, -1.2) {};
  \node[font=\tiny, anchor=west] at (1.0, -1.2) {Trainable};

\end{scope}

\node[font=\scriptsize, text=gray!70] at (0, 6.7)
  {\textit{Round $t$:\; distribute\;
   $\bar{\theta}^{(t)}$ \;$\rightarrow$\; local train \;$\rightarrow$\;
   upload\; $\Delta\theta_k$ \;$\rightarrow$\; aggregate}};

\end{tikzpicture}
\caption{Federated LoRA framework for RF interference mitigation within an
O-RAN deployment. The Near-RT RIC serves as the aggregation server,
coordinating federated learning across $K$ distributed gNBs. Each gNB trains
lightweight LoRA adapters on its local interference data while the WaveNet
backbone remains frozen. Right: LoRA placement on the dilated convolutional
layer within each residual block.}
\label{fig:system}
\end{figure*}

\subsection{System Architecture}
We consider a disaggregated O-RAN deployment consisting of $K$ gNBs, each     
comprising an O-RU for RF front-end processing and an O-DU for baseband
computation. A Near-RT RIC serves as the central coordination point, connected
to all gNBs via bandwidth-limited midhaul/backhaul interfaces. The signal separation model executes at the gNB, operating directly on time-domain IQ baseband samples. The Near-RT RIC acts as the
federated aggregation server, collecting adapter parameter updates from
participating or clustered gNBs and distributing the aggregated global adapter each communication round. This deployment is illustrated in Fig. 2.

Each gNB $k \in \{1, \dots, K\}$ operates in a distinct RF environment
characterized by a local interference profile. In practice, a gNB near an
industrial facility may experience electromagnetic interference (EMI), while
another in a dense urban area may encounter co-channel communication signals.
This spatial heterogeneity produces a naturally non-IID data distribution
across the network, where the non-IID structure arises from physically
distinct interference environments rather than artificial label skew.

\subsection{Signal Model}

At each gNB $k$, the received baseband signal is modeled as:
\begin{equation}
  y_k(t) = s(t) + i_k(t), \quad t = 1, \dots, T,
  \label{eq:mixture}
\end{equation}
where $s(t) \in \mathbb{C}$ is the signal of interest (SOI) and $i_k(t) \in
\mathbb{C}$ is the node-specific interference. Following the RF Challenge
convention~\cite{lancho2025rfchallenge}, we use signal-to-interference-plus-noise ratio (SINR) as the interference signals are derived from recorded RF captures that inherently contain additive noise. Each interference signal is scaled to achieve a target SINR and undergoes a random phase rotation before being added to the SOI. During training, the SINR is drawn uniformly at random from $[-10, +10]$~dB; at test time, performance is evaluated at 11 discrete levels from $-10$ to $+10$~dB in 2~dB increments.
The SOI is an OFDM signal with QPSK modulation, consisting of $T = 40{,}960$ complex-valued samples at a 20~MHz sampling rate. Each signal spans 512 OFDM symbols carrying $B = 57{,}344$ uncoded information bits.

The interference $i_k(t)$ is drawn from a node-specific distribution
$\mathcal{I}_k$, which may include communication signals (CommSignal2,
CommSignal3), electromagnetic interference (EMISignal1), or mixtures thereof.
The SINR is defined as the ratio of SOI power to interference-plus-noise power:
\begin{equation}
  \text{SINR} = 10 \log_{10} \frac{P_s}{P_i} \quad [\text{dB}],
  \label{eq:sinr}
\end{equation}
where $P_s = \frac{1}{T}\|s\|^2$ and $P_i = \frac{1}{T}\|i_k\|^2$ denote the
average signal and interference power, respectively. 

The objective is to recover an estimate $\hat{s}(t)$ from the mixture
$y_k(t)$, from which information bits are obtained via OFDM demodulation and
QPSK demapping. Performance is measured by the bit error rate (BER) between
the estimated and true bit sequences.

\subsection{Problem Formulation}

Let $f_{W, \theta}$ denote the signal separation model parameterized by
backbone weights $W$ and adapter parameters $\theta$. The model maps a mixture
input to an estimated SOI: $\hat{s} = f_{W, \theta}(y)$. The local objective
at gNB $k$ is to minimize the mean squared error between the estimated and
true SOI over its local dataset $\mathcal{D}_k$:
\begin{equation}
  \mathcal{L}_k(\theta) = \frac{1}{|\mathcal{D}_k|} \sum_{(y, s) \in
\mathcal{D}_k} \| f_{W, \theta}(y) - s \|^2,
  \label{eq:local_loss}
\end{equation}
where the backbone $W$ is frozen after centralized pretraining and only the
adapter parameters $\theta$ are optimized. This separation reflects the
assumption that the backbone has learned general signal extraction
capabilities from diverse training data, while the adapters capture
node-specific interference characteristics.

\subsection{Federated Adapter Aggregation}

In the federated setting, the global objective is to find adapter parameters
that minimize the weighted average of local losses across all $K$ gNBs:
\begin{equation}
  \min_{\theta} \; \mathcal{L}(\theta) = \sum_{k=1}^{K}
\frac{|\mathcal{D}_k|}{\sum_{j} |\mathcal{D}_j|} \, \mathcal{L}_k(\theta).
  \label{eq:global_obj}
\end{equation}

We solve this via federated averaging (FedAvg)~\cite{mcmahan2017communication}
restricted to the adapter parameters $\theta$. At each communication round
$t$:

\begin{enumerate}
  \item The aggregation server distributes the current global adapter
$\bar{\theta}^{(t)}$ to all $K$ gNBs.
  \item Each gNB $k$ initializes its local adapter from $\bar{\theta}^{(t)}$
and performs $E$ epochs of local SGD on $\mathcal{D}_k$, producing updated
parameters $\theta_k^{(t)}$.
  \item The server aggregates local adapters via weighted averaging:
  \begin{equation}
      \bar{\theta}^{(t+1)} = \sum_{k=1}^{K} \frac{|\mathcal{D}_k|}{\sum_{j}
|\mathcal{D}_j|} \, \theta_k^{(t)}.
      \label{eq:fedavg}
  \end{equation}
\end{enumerate}

Crucially, only $\theta$ is transmitted each round, not the full model $\{W,
\theta\}$. When $\theta$ consists of LoRA adapter matrices, the communication
cost per round per node is $|\theta|$, compared to $|W| + |\theta|$ for
standard FedAvg over the full model. In our architecture, $|\theta| =
14{,}400$ at rank=4 while $|W| = 281{,}954$, yielding a ${\sim}20\times$ reduction in
per-round communication.

After local training at round $t$, each gNB retains its locally adapted model
$f_{W, \theta_k^{(t)}}$ for inference, while contributing $\theta_k^{(t)}$ to
the global aggregate. This structure implicitly provides personalization: the
frozen backbone encodes shared signal structure learned during centralized
pretraining, while each node's locally trained adapter captures its
site-specific interference characteristics. The global aggregate
$\bar{\theta}^{(t+1)}$ seeds the next round of local adaptation, enabling
knowledge transfer across nodes --- particularly benefiting data-scarce nodes
that have limited local exposure to certain interference types.

\subsection{End-to-End Workflow}
The complete workflow proceeds in two phases. In the \textit{pretraining
phase}, a WaveNet backbone is trained centrally on a large corpus of signal
mixtures spanning multiple interference types. This backbone learns
general-purpose signal extraction capabilities and is subsequently distributed
to all $K$ gNBs as a frozen feature extractor.

In the \textit{federated adaptation phase}, each gNB augments the frozen
backbone with lightweight LoRA adapters and participates in $R$ rounds of
federated learning. At each round, gNBs perform local adapter training on
their private interference data for $E$ epochs, upload the updated adapter
parameters to the Near-RT RIC, and receive the aggregated global adapter to
seed the next round. After adaptation, each gNB retains its locally trained
adapter for real-time inference on incoming signal mixtures, applying OFDM
demodulation to the separated signal to recover information bits. The entire
adaptation phase communicates only adapter parameters ($|\theta| \ll |W|$),
making it feasible over bandwidth-limited backhaul links.
\section{Adapter Design for Dilated Convolutional Networks}  
\subsection{WaveNet Architecture}
The backbone signal separation model is based on the WaveNet                  
architecture~\cite{oord2016wavenet}, originally proposed for speech generation
and subsequently adapted for speech denoising~\cite{rethage2018wavenet}. The
model operates directly on time-domain IQ samples, taking the real and
imaginary components of the mixture signal $y_k(t)$ as a 2-channel input and
producing a 2-channel output representing the estimated SOI $\hat{s}(t)$.

The architecture consists of an input projection, a stack of $R = 15$ residual
blocks, and an output projection. The input projection is a $1 \times 1$
convolution that maps the 2-channel input to $C = 48$ residual channels. Each
residual block $i$ contains two convolution layers:

\begin{itemize}
  \item A \textit{dilated convolution} (kernel size 3, dilation $d_i = 2^{i
\bmod m}$ with cycle length $m = 5$) that maps $C \to 2C$ channels, performing
temporal filtering at an exponentially growing receptive field scale.
  \item A $1 \times 1$ \textit{output projection} that maps $C \to 2C$
channels, performing channel mixing.
\end{itemize}

The $2C$ output of the dilated convolution is split into two halves and passed
through a gated activation:
\begin{equation}
  \mathbf{h}_i = \sigma(\mathbf{g}_i) \odot \tanh(\mathbf{v}_i),
  \label{eq:gated_activation}
\end{equation}
where $\mathbf{g}_i$ and $\mathbf{v}_i$ are the gate and filter components
obtained by splitting the dilated convolution output along the channel
dimension. The activated output is then passed through the $1 \times 1$
projection and split into a residual connection and a skip connection:
\begin{equation}
  \mathbf{x}_{i+1} = \frac{\mathbf{x}_i + \mathbf{r}_i}{\sqrt{2}}, \quad
\mathbf{s}_i = \text{skip}_i,
  \label{eq:residual}
\end{equation}
where $\mathbf{r}_i$ and $\mathbf{s}_i$ are the residual and skip components.
The skip connections from all $R$ blocks are summed, normalized by
$1/\sqrt{R}$, and passed through a final skip projection, ReLU activation, and
output projection to produce the 2-channel estimated SOI.

The dilation cycle $\{1, 2, 4, 8, 16\}$ repeats three times across the 15
blocks, giving the model multi-scale temporal receptive fields ranging from 3
to 33 samples per cycle. The full lightweight model contains 281,954 parameters and was
pretrained centrally for 151,200 steps on 56,000 signal mixtures (28,000
CommSignal2 + 28,000 CommSignal3) using Adam optimization. Notably, the
pretraining data contains no EMI samples, so adaptation to EMI-containing
nodes relies entirely on the PEFT methods.

\subsection{LoRA on Dilated Convolutions}

Each WaveNet residual block contains two convolution layers: a dilated      
convolution for temporal filtering and a $1 \times 1$ projection for channel  
mixing. We apply LoRA to the dilated convolutions rather than the $1 \times 1$
projections. This design choice is motivated by the role each layer plays in 
signal separation: dilated convolutions capture temporal interference patterns
at multiple time scales, while $1 \times 1$ convolutions mix channel         
representations. Since interference mitigation requires adapting the model's
temporal filtering behavior to site-specific interference characteristics,
targeting the dilated convolutions allows the adapter to directly modify the
temporal receptive field response at each scale.

We add a parallel low-rank branch to the dilated convolution in each residual
block $i$:
\begin{equation}
  \mathbf{y}_i = \text{DilConv}_i(\mathbf{x}) + \frac{\alpha}{r}
\mathbf{B}_i(\mathbf{A}_i(\mathbf{x})),
  \label{eq:lora}
\end{equation} 
where $\mathbf{A}_i \in \text{Conv1d}$ is a down-projection with $C \to r$
  channels, kernel size 3, and dilation $d_i$; and $\mathbf{B}_i \in
  \text{Conv1d}$ is a pointwise up-projection with $r \to 2C$ channels and
  kernel size 1.
The scaling factor $\alpha / r$ controls the magnitude of the adapter's
contribution. Critically, $\mathbf{A}_i$ preserves the kernel size and
dilation rate $d_i = 2^{i \bmod m}$ of the original dilated convolution,
ensuring the low-rank branch operates over the same temporal receptive field
as the frozen layer it augments.

Following standard LoRA initialization~\cite{hu2022lora}, $\mathbf{A}_i$ is
initialized with Kaiming uniform and $\mathbf{B}_i$ is initialized to zeros,
so that $\mathbf{B}_i(\mathbf{A}_i(\mathbf{x})) = 0$ at the start of training.
This preserves the pretrained backbone's behavior before any adaptation
occurs.

With rank $r = 4$ and $C = 48$ channels, each block contributes $r \times C
\times 3 + 2C \times r \times 1 = 576 + 384 = 960$ parameters for
$\mathbf{A}_i$ and $\mathbf{B}_i$ respectively. Across all $R = 15$ blocks,
the total adapter size is $15 \times 960 = 14{,}400$ trainable parameters,
representing 5.1\% of the 281,954 backbone parameters. This compact
parameterization is what enables efficient federated communication --- each
round requires transmitting only 14,400 parameters per node rather than the
full model, yielding a ${\sim}20\times$ reduction in per-round communication
cost.

We evaluate ranks $r \in \{2, 4, 8\}$ and find diminishing returns beyond $r =
4$: increasing to $r = 8$ doubles the adapter size but yields only 0.9
percentage points of additional BER improvement, confirming $r = 4$ as an
efficient operating point (see Fig.~\ref{fig:comm_tradeoff}).

\subsection{FiLM Conditioning}

As a lightweight baseline, we additionally evaluate Feature-wise Linear       
Modulation (FiLM)~\cite{perez2018film}, which applies a learned channel-wise
affine transformation to the residual stream between successive blocks:       
\begin{equation}
  \mathbf{x}'_i = \boldsymbol{\gamma}_i \odot \mathbf{x}_i +
\boldsymbol{\beta}_i,
  \label{eq:film}
\end{equation}
where $\boldsymbol{\gamma}_i, \boldsymbol{\beta}_i \in \mathbb{R}^{C}$ are
per-block scale and shift parameters and $\odot$ denotes channel-wise
multiplication. Parameters are initialized to $\boldsymbol{\gamma}_i =
\mathbf{1}$, $\boldsymbol{\beta}_i = \mathbf{0}$ (identity), preserving the
pretrained backbone at initialization. With $R = 15$ blocks and $C = 48$
channels, FiLM adds only $2 \times R \times C = 1{,}440$ trainable parameters
(0.51\% of the backbone).

Unlike LoRA, which introduces new temporal filter responses via low-rank
convolutions on the dilated layers, FiLM can only rescale and shift existing
channel activations. This distinction is important: FiLM modulates
\textit{what the backbone already computes}, while LoRA augments \textit{how
the backbone filters in time}. As we show in Section~\ref{sec:results}, this
limits FiLM's effectiveness for interference suppression, where adapting
temporal filtering behavior is critical.

\section{Experimental Setup}
\subsection{Dataset and Node Configuration}

Signal mixtures are generated following the ICASSP 2024 RF Signal Separation
Challenge protocol~\cite{lancho2025rfchallenge}. Three interference types are
considered: CommSignal2, CommSignal3, and EMISignal1, drawn from recorded RF
captures in the challenge dataset.

Training data is distributed across $K = 5$ gNBs in a non-IID manner
reflecting heterogeneous interference environments. Nodes 1 and 2 encounter
only communication interference (CommSignal2 and CommSignal3, respectively),
Nodes 3 and 4 face mixed communication and electromagnetic interference, and
Node 5 encounters only EMISignal1. Crucially, the centrally pretrained
backbone has no EMI exposure, so adaptation to EMI-containing nodes relies
entirely on the PEFT methods.

We evaluate two non-IID data regimes:
\begin{itemize}
  \item \textbf{Balanced:} Each node has 3,000 training samples drawn from
its local interference distribution. Nodes differ in interference type but not
data volume.
  \item \textbf{Imbalanced:} EMI samples are reduced to 200 per node (from
1,000 or 3,000) while communication-interference samples remain unchanged,
simulating rare interference exposure at the edge.
\end{itemize}

Test sets are constructed at 11 discrete SINR levels from $-10$ to $+10$~dB in
2~dB increments. Each node is evaluated on a local test set matching its
interference profile, as well as a global test set containing all three
interference types.

\subsection{Model Training}

The WaveNet backbone was pretrained centrally for 151,200 steps on 56,000
signal mixtures (28,000 CommSignal2 + 28,000 CommSignal3) using Adam with a
learning rate of $5 \times 10^{-4}$ and batch size 8 with FP16 mixed
precision. The pretrained backbone serves as the frozen feature extractor for
all subsequent adaptation methods.

We compare six adaptation strategies: (1)~\textbf{Backbone}: the frozen
pretrained model with no adaptation; (2)~\textbf{FedAvg}: standard federated
averaging over all 281,954 model parameters; (3)~\textbf{L-FiLM} and
(4)~\textbf{Fed-FiLM}: local and federated FiLM conditioning (1,440 params);
(5)~\textbf{L-LoRA} and (6)~\textbf{Fed-LoRA}: local and federated LoRA on
dilated convolutions (14,400 params); and (7)~\textbf{Full-FT}: local full
fine-tuning of all parameters as an upper bound.

For local adaptation (L-LoRA, L-FiLM, Full-FT), each node trains for 20 epochs
with a ReduceLROnPlateau scheduler and early stopping based on validation
loss. For federated methods (FedAvg, Fed-LoRA, Fed-FiLM), training proceeds
for $R = 10$ communication rounds with $E = 2$ local epochs per round per
node. All methods use a 90/10 train/validation split.
Table~\ref{tab:training_config} summarizes the key hyperparameters.

\begin{table}[t]
\centering
\caption{Training configuration.}
\label{tab:training_config}
\begin{tabular}{|l|c|}
\hline
\textbf{Parameter} & \textbf{Value} \\
\hline
\multicolumn{2}{|c|}{\textit{Backbone Pretraining}} \\
\hline
Training steps & 151,200 \\
Training data & 56,000 (28k CS2 + 28k CS3) \\
Learning rate & $5 \times 10^{-4}$ \\
\hline
\multicolumn{2}{|c|}{\textit{Local Adaptation}} \\
\hline
Epochs & 20 \\
Scheduler & ReduceLROnPlateau \\
\hline
\multicolumn{2}{|c|}{\textit{Federated Adaptation}} \\
\hline
Communication rounds $R$ & 10 \\
Local epochs $E$ & 2 \\
Aggregation & Weighted FedAvg \\
\hline
\multicolumn{2}{|c|}{\textit{Shared}} \\
\hline
Batch size & 8 \\
Optimizer & Adam \\
LR (FedAvg / Full-FT) & $10^{-4}$ \\
LR (LoRA / FiLM) & $10^{-3}$ \\
LoRA rank $r$ & 4 \\
FP16 & \checkmark \\
Train/val split & 90/10 \\
\hline
\multicolumn{2}{|c|}{\textit{Params per round per node}} \\
\hline
FedAvg & 281,954 \\
Fed-LoRA ($r\!=\!4$) & 14,400 \\
Fed-FiLM & 1,440 \\
\hline
\end{tabular}
\end{table}

\begin{table*}[t]                                                             
\centering                                                                    
\caption{Average BER across all SINR levels for each adaptation method under balanced dataset. Full-FT fine-tunes all 281{,}954 parameters; L-LoRA adapts only 14{,}400. Bold = best adapter method (excluding Full-FT upper bound).}    
\label{tab:full_data_results}
\begin{tabular}{|l|c|c|c|c|c|c|c|}
\hline
\textbf{Node} & \textbf{Backbone} & \textbf{FedAvg} & \textbf{L-FiLM} &
\textbf{Fed-FiLM} & \textbf{L-LoRA} & \textbf{Fed-LoRA} & \textbf{Full-FT} \\
\hline
1 (CS2)     & .00447 & .01128 & .00418 & .00420 & \textbf{.00411} & .00414 &
.00399 \\
\hline
2 (CS3)     & .12596 & .12716 & .12596 & .12596 & \textbf{.12586} & .12591 &
.12558 \\
\hline
3 (CS2+EMI) & .01397 & .01142 & .01131 & .01175 & \textbf{.00670} & .00706 &
.00569 \\
\hline
4 (CS3+EMI) & .07476 & .06975 & .07252 & .07311 & \textbf{.06783} & .06828 &
.06702 \\
\hline
5 (EMI)     & .02319 & .01157 & .01304 & .01303 & .00672 & \textbf{.00653} &
.00554 \\
\hline
\textbf{Avg} & .04847 & .04624 & .04540 & .04561 & \textbf{.04225} & .04238 &
.04156 \\
\hline
\end{tabular}
\end{table*}

\begin{table*}[t]
\centering
\caption{Average BER under data scarcity (imbalanced regime).}
\label{tab:scarce_results}
\begin{tabular}{|l|c|c|c|c|c|}
\hline
\textbf{Node} & \textbf{Backbone} & \textbf{FedAvg} & \textbf{L-LoRA} &
\textbf{Fed-LoRA} & \textbf{Full-FT} \\
\hline
1 (CS2)     & .00447 & .01068 & \textbf{.00411} & .00416 & .00397 \\
\hline
2 (CS3)     & .12596 & .12627 & \textbf{.12589} & .12593 & .12558 \\
\hline
3 (CS2+EMI) & .01397 & .01357 & \textbf{.00814} & .00863 & .00720 \\
\hline
4 (CS3+EMI) & .07476 & .07167 & \textbf{.06856} & .06970 & .06800 \\
\hline
5 (EMI)     & .02319 & .01632 & .01335 & \textbf{.01237} & .01229 \\
\hline
\textbf{Avg} & .04847 & .04770 & \textbf{.04401} & .04416 & .04341 \\
\hline
\end{tabular}
\end{table*}

\section{Results and Analysis}
\label{sec:results}
\begin{figure}[t]                                                                                                                                                                                              
\centering      
\begin{tikzpicture}                                                                                                                                                                                            
\begin{semilogxaxis}[                                                                                                                                                                                          
  width=\columnwidth,                                                                                                                                                                                        
  height=0.7\columnwidth,        
  xlabel={Parameters communicated per round},
  ylabel={Avg.\ BER improvement (\%)},
  xmin=800, xmax=500000,
  ymin=0, ymax=16,
  grid=both,
  grid style={gray!20},
  legend style={at={(0.03,0.97)}, anchor=north west, font=\small},
  every axis label/.style={font=\small},
  tick label style={font=\footnotesize},
]

\addplot[
  only marks, mark=triangle*, mark size=4pt,
  color=orange!80!black, fill=orange!60,
] coordinates {(1440, 5.9)};
\addlegendentry{Fed-FiLM}

\addplot[
  only marks, mark=*, mark size=4pt,
  color=blue!80!black, fill=blue!60,
] coordinates {(7200, 11.3) (14400, 12.6) (28800, 13.5)};
\addlegendentry{Fed-LoRA}

\addplot[
  only marks, mark=square*, mark size=4pt,
  color=red!80!black, fill=red!60,
] coordinates {(281954, 4.6)};
\addlegendentry{FedAvg}

\node[anchor=south, font=\scriptsize] at (axis cs:1440, 5.9) {1.4k};
\node[anchor=south west, font=\scriptsize] at (axis cs:7200, 11.3) {$r\!=\!2$};
\node[anchor=south, font=\scriptsize] at (axis cs:14400, 12.6) {$r\!=\!4$};
\node[anchor=south west, font=\scriptsize] at (axis cs:28800, 13.5) {$r\!=\!8$};
\node[anchor=south, font=\scriptsize] at (axis cs:281954, 4.6) {282k};

\addplot[dashed, thin, blue!50, no markers] coordinates {(7200, 11.3) (14400, 12.6) (28800, 13.5)};

\end{semilogxaxis}
\end{tikzpicture}
\caption{Communication--performance tradeoff for federated methods.}
\label{fig:comm_tradeoff}
\end{figure}
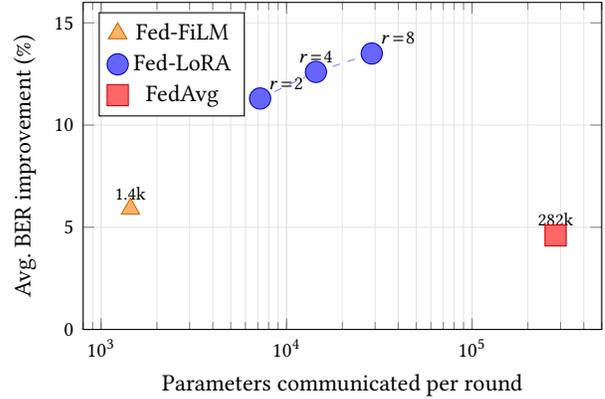

\subsection{Balanced Data Regime}                                             
                  
Table \ref{tab:full_data_results} reports the average BER across all 11 SINR levels
for each adaptation method evaluated on per-node local test sets.
Several findings emerge.

\textbf{Local LoRA achieves the best average BER among all lightweight
methods}, reducing the backbone's average BER by 12.8\% while training only
5.1\% of the model parameters (14,400 vs.\ 281,954).
Fed-LoRA performs comparably at 12.6\%, confirming that federated aggregation
of adapter parameters preserves most of the local adaptation quality.
Both LoRA variants substantially outperform FiLM-based methods (L-FiLM: 6.3\%,
Fed-FiLM: 5.9\%), which lack the capacity to synthesize new filter responses
and are limited to rescaling existing features.

\textbf{Full fine-tuning provides a modest upper bound.}
Local full fine-tuning achieves 14.3\% average improvement, meaning L-LoRA
captures approximately 90\% of the full fine-tuning gain with 20$\times$ fewer
trainable parameters.
The gap is largest on EMI-heavy nodes (Node~3: 59.3\% vs.\ 52.0\%; Node~5:
76.1\% vs.\ 71.0\%), suggesting that EMI suppression benefits from additional
model capacity, though LoRA still captures the majority of the gain.

\textbf{FedAvg degrades performance on well-represented nodes.}
While FedAvg achieves a reasonable 4.6\% average improvement, this masks a
harmful trade-off: it improves Node~5 (EMI) by 50.1\% but degrades Node~1
(CS2) by 152.6\%.
Aggregating all 281,954 backbone parameters causes catastrophic interference
between the heterogeneous node objectives.
In contrast, Fed-LoRA improves every node without degrading any, demonstrating
that restricting federation to a small adapter subspace avoids this failure
mode.

\textbf{CommSignal3 remains a hard interference type.}
Node~2 (CS3-only) shows negligible improvement across all methods, including
full fine-tuning (+0.3\%).
This suggests that CS3 interference is structurally difficult to suppress
given the current backbone architecture, and the bottleneck lies in the
model's representational capacity rather than insufficient adaptation.

\begin{figure*}[t]                                                                            
\centering
\includegraphics[width=\textwidth]{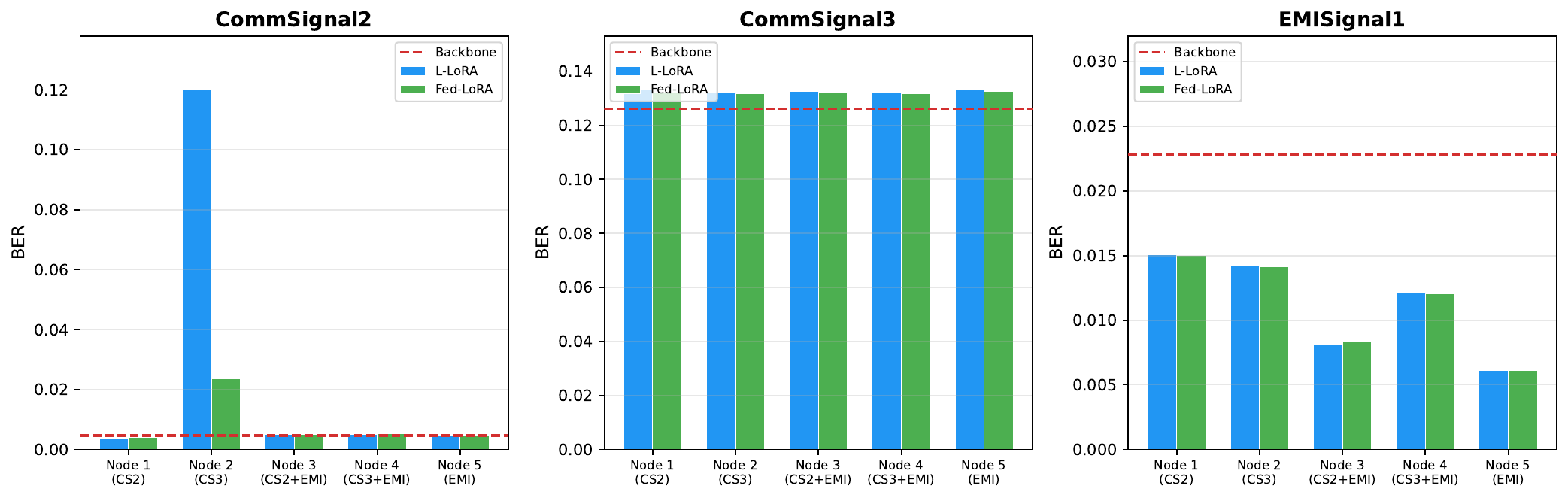}                               
\caption{Per-interference-type BER on the global test set}
\label{fig:per_type_ber}
\end{figure*}

\subsection{Imbalanced Data Regime}

To evaluate robustness under data scarcity, we reduce EMI training samples to
200 per node (from 3,000) while keeping CS2 and CS3 volumes unchanged.
This simulates a realistic deployment scenario where certain interference
types are rarely observed.
Table~\ref{tab:scarce_results} reports the results.

\textbf{L-LoRA remains the strongest method overall} (9.2\% average
improvement), but the gap with Fed-LoRA narrows compared to the balanced
regime (0.3pp vs.\ 0.2pp).
Both methods degrade gracefully: L-LoRA's Node~3 improvement drops from 52.0\%
to 41.8\%, while Fed-LoRA drops from 49.5\% to 38.2\%.

\textbf{Federation provides a clear advantage on the most data-starved node.}
On Node~5 (200 EMI-only samples), Fed-LoRA achieves 46.6\% improvement versus
L-LoRA's 42.5\%---a 4.1 percentage point gap.
This is the only node where Fed-LoRA consistently outperforms L-LoRA, and the
advantage is attributable to knowledge transfer: the federated aggregate
incorporates EMI-relevant gradient information from Nodes~3 and~4, effectively
augmenting Node~5's limited local data.

\textbf{FedAvg is catastrophic under scarcity.}
FedAvg degrades Node~1 by 139.1\%, consistent with the catastrophic
~degradation observed in the balanced regime (-152.6\%).
Its average improvement of just 1.6\% makes it unsuitable for heterogeneous
deployments.

\textbf{Full fine-tuning under scarcity.}
Local full fine-tuning achieves 10.5\% average improvement, outperforming                     
all other methods. On Node~5, Fed-LoRA (46.6\%) nearly matches full
fine-tuning (47.0\%) despite training 20$\times$ fewer parameters, showing
that federated knowledge transfer can largely compensate for limited local
model capacity under data scarcity.

\vspace{-6pt} 
\subsection{Per-Interference-Type Analysis}

Figure~\ref{fig:per_type_ber} disaggregates BER by interference type on the
global test set, revealing how each node's locally adapted model generalizes
across interference types it may not have trained on.

\textbf{CommSignal2:} Most nodes achieve near-backbone BER, except Node~2
(trained only on CS3), which exhibits 0.120 BER under L-LoRA.
Fed-LoRA reduces Node~2's CS2 BER to 0.024---an 80\% reduction---demonstrating
that federation transfers CS2 suppression knowledge from other nodes.
This is the clearest example of cross-node knowledge transfer in our
experiments.

\textbf{CommSignal3:} All nodes cluster near the backbone BER ($\sim$0.13),
confirming that CS3 is uniformly difficult and neither local nor federated
adaptation provides meaningful improvement on this interference type.

\textbf{EMISignal1:} Both L-LoRA and Fed-LoRA reduce BER well below the
backbone across all nodes.
The improvement is largest on nodes that include EMI in their training data
(Nodes~3, 4, 5), but even CS-only nodes (1, 2) show substantial EMI
suppression, indicating that the backbone already captures partial EMI
structure and LoRA refines it.

\subsection{Communication--Performance Trade-off}

Fig.~\ref{fig:comm_tradeoff} plots average BER improvement plots average BER improvement against per-round
communication cost (parameters transmitted per node) for all federated methods
across ranks $r \in \{2, 4, 8\}$.

\begin{figure*}[t]                                                                            
\centering
\includegraphics[width=\textwidth]{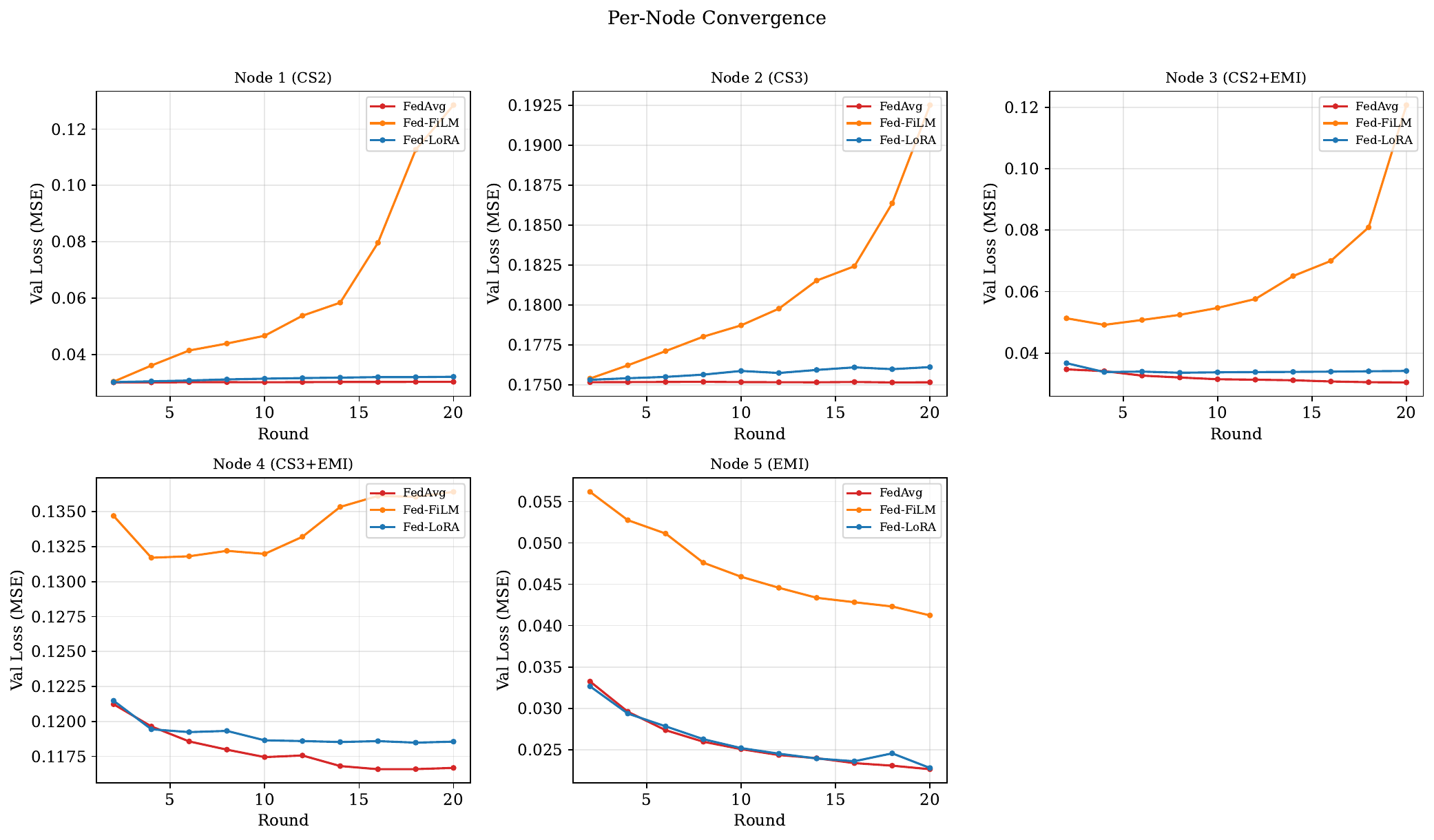}                                  
\caption{Per-node validation loss (MSE) across federated communication rounds.}
\vspace{-6pt}
\label{fig:convergence}
\end{figure*}

Fed-LoRA achieves 11.3\% improvement at rank~2 (7,200 params/node/round),
increasing to 12.6\% at rank~4 (14,400) and 13.5\% at rank~8 (28,800).
This represents a favorable trade-off: Fed-LoRA at rank~2 already outperforms
both FedAvg (4.6\% at 281,954 params) and Fed-FiLM (5.9\% at 1,440 params)
while transmitting 39$\times$ fewer parameters than FedAvg.

Diminishing returns are evident: doubling the rank from 4 to 8 yields only
0.9pp additional improvement for Fed-LoRA.
L-LoRA follows the same trend (11.7\% $\rightarrow$ 12.8\% $\rightarrow$
13.7\%), consistently outperforming Fed-LoRA by 0.2--0.4pp at each rank,
confirming that the federation overhead is small but nonzero.

FedAvg occupies an unfavorable position: it transmits 10--39$\times$ more
parameters than Fed-LoRA while achieving 2--3$\times$ less improvement, making
it Pareto-dominated across the entire rank sweep.

\vspace{-6pt} 
\subsection{Convergence Analysis}

Fig.~\ref{fig:convergence} shows per-node validation loss across federated
communication rounds.

\textbf{Fed-FiLM diverges under federation.} Fed-FiLM's validation loss
increases monotonically after the first 1--2 rounds on four of five nodes,
with only Node~5 (EMI) showing sustained improvement. With only 1,440
scale-and-shift parameters, FiLM lacks sufficient capacity to reconcile
conflicting adaptation directions from heterogeneous nodes during
aggregation.

\textbf{Fed-LoRA converges stably.} Fed-LoRA stabilizes by round~3--4
across all nodes with no divergence, confirming that 14,400 low-rank
parameters provide enough expressiveness for stable federated aggregation.

\textbf{MSE--BER disconnect in FedAvg.} FedAvg converges to the lowest
validation loss on most nodes, yet produces the worst BER on Node~1. Minimizing aggregate MSE across
heterogeneous interference types does not guarantee correct signal recovery for demodulation at individual nodes.

\vspace{-6pt} 
\section{Conclusion and Future Work}

This paper investigated parameter-efficient adaptation strategies for RF      
interference mitigation in federated wireless deployments. We proposed
applying LoRA to dilated convolutional layers of a WaveNet-based signal       
separation backbone, enabling node-specific adaptation with only 5.1\% of the
backbone parameters. We evaluated six adaptation strategies---local and
federated variants of LoRA, FiLM, and full fine-tuning---across balanced and
imbalanced data regimes on a 5-node heterogeneous interference scenario.

Our results demonstrate that L-LoRA is the strongest lightweight
method overall, capturing approximately 90\% of full fine-tuning's gain while
training 20$\times$ fewer parameters. Fed-LoRA performs
comparably in most settings and provides a measurable advantage on
data-starved nodes, where cross-node knowledge transfer compensates for
limited local observations. In contrast, standard FedAvg over the full
backbone suffers catastrophic degradation on well-represented nodes,
confirming that restricting federation to a compact adapter subspace is
essential for heterogeneous interference environments.

The per-interference-type analysis further reveals that federation enables
cross-node generalization: Fed-LoRA reduces a CS3-trained node's CommSignal2
BER by 80\% compared to L-LoRA, demonstrating effective transfer of
interference suppression capabilities across nodes with non-overlapping data
distributions. The communication--performance trade-off analysis shows that
Fed-LoRA at rank~2 already outperforms FedAvg while transmitting 39$\times$
fewer parameters per round, making it practical for bandwidth-constrained
fronthaul links. Convergence analysis further reveals that na\"ive federation of lightweight adapters (Fed-FiLM) can diverge under non-IID interference distributions, whereas Fed-LoRA maintains stable convergence across all nodes.

Several directions remain for future investigation.
First, our evaluation uses a simulated federated setting with sequential node
training. Real-world deployment and testing on a real O-RAN testbed with asynchronous communication and Near-RT RIC orchestration would validate the practical feasibility of federated adapter aggregation.
Second, CommSignal3 proved resistant to all adaptation methods including full fine-tuning, suggesting that architectural modifications to the backbone may be necessary, such as an increased receptive field or attention mechanisms.
Third, adaptive rank allocation across nodes or layers could improve
efficiency. Nodes with simple interference may require lower
rank than mixed-interference nodes, and deeper layers may benefit from higher
rank than shallow layers.
Finally, extending this framework to additional PEFT methods such as adapter
layers or prompt tuning, and to other PHY-layer tasks beyond signal separation
(e.g., channel estimation, equalization), would broaden the applicability of
federated parameter-efficient adaptation at the wireless edge.

\begin{acks}                                                                                  
This research was supported by the U.S. Department of Commerce’s National Telecommunications and Information Administration (NTIA) under the Public Wireless Supply Chain Innovation Fund Grant Program (Award 24-60-IF2415: ASPEN - Advanced Signal Processing Enhancement for Next-Generation Open Radio Units), administered by the National Institute of Standards and Technology.
\end{acks}
\bibliographystyle{ACM-Reference-Format}
\bibliography{references}

\end{document}